# Coherent interaction between free electrons and cavity photons

*Kangpeng Wang[1], Raphael Dahan[1], Michael Shentcis[1], Yaron Kauffmann[2], Shai Tsesses[1], and Ido Kaminer[1] \**

1. Solid State Institute and Faculty of Electrical Engineering, Technion – Israel Institute of Technology, Haifa 32000, Israel
2. Department of Materials Science & Engineering, Technion – Israel Institute of Technology, Haifa 32000, Israel

\* *kaminer@technion.ac.il*

**Abstract**

Since its inception, research of cavity quantum electrodynamics (CQED)[1,2] has greatly extended our understanding of light–matter interactions and our ability to utilize them. Thus far, all the work in this field has been focused on light interacting with bound electron systems – such as atoms, molecules, quantum dots, and quantum circuits. In contrast, markedly different physical phenomena and applications could be found in free-electron systems, the energy distribution of which is continuous and not discrete, implying tunable transitions and selection rules. In addition to their uses for electron microscopy[3,4], the interaction of free electrons with light gives rise to important phenomena such as Cherenkov radiation, Compton scattering, and free-electron lasing[5,6]. Advances in the research of ultrafast electron–light interactions have also enabled the development of powerful tools for exploring femtosecond dynamics at the nanoscale[7,8]. However, thus far, no experiment has shown the integration of free electrons into the framework of CQED, because the fundamental electron–light interaction is limited in strength and lifetime. This limit explains why many phenomena have remained out of reach for experiments with free electrons. In this work, we developed the platform for studying CQED at the nanoscale with free electrons and demonstrated it by observing their coherent interaction with cavity photons for the first time. To demonstrate this concept, we directly measure the cavity photon lifetime via a free electron probe and show more than an order of magnitude enhancement in the electron–photon interaction strength. These capabilities may open new paths toward using free electrons as carriers of quantum information, even more so after strong coupling between free electrons and cavity photons will have been demonstrated. Efficient electron–cavity photon coupling could also allow new nonlinear phenomena of cavity opto-electro-mechanics and the ultrafast exploration of soft matter or other beam-sensitive materials using low electron current and low laser exposure.

**Introduction**

The ultrafast interaction between free electrons and laser pulses, mediated by evanescent light fields, has enabled the development of powerful tools to explore femtosecond dynamics at the nanoscale[9–11]. This type of interaction has been used to demonstrate a laser-driven quantum walk of free electrons[12], attosecond electron pulse trains[13–15], the transfer of angular momentum from the optical nearfield to free electrons[16], and the imaging of plasmons[9,17,18] at (laser-induced) meV energy resolution[10]. However, all coherent electron–photon interactions observed thus far were fundamentally limited in their interaction strength and lifetime because of losses and/or low quality-factors.

In this study, we integrated free electrons into a nanoscale CQED platform and observed the effects of coherent electron–light interaction created by trapping the light in photonic cavities[20] (Fig. 1). In such cavities, we demonstrated the quantization of the electron's kinetic energy in record-low pulse energies. We also directly measured, for the first time, the lifetime of photons in a cavity via a free electron probe. This prolonged lifetime and the order-of-magnitude decrease in the required pulse energies, both show the *enhancement of the intrinsic electron–photon interaction* provided by the photonic cavity. Additionally, we are able to resolve the photonic bandstructure as a function of energy, momentum, and polarization, simultaneously capturing the spatial distribution of the cavity modes at deep-subwavelength resolution. These capabilities offer multidimensional characterization of nanoscale optically excitable systems, going beyond the limits of incoherent broadening, which is particularly attractive in systems such as quantum dots[21], pico-cavities[22] and van-der-Waals heterostructures[23,24]. Our results may open new pathways toward studying nanoscale CQED with free electrons, the implications of which have been discussed in several recent theory papers that predicted strong coupling of free electrons with cavity photons[25,26], electron–photon and electron–electron entanglement[25,27], and new physics in Cherenkov radiation[28–30].

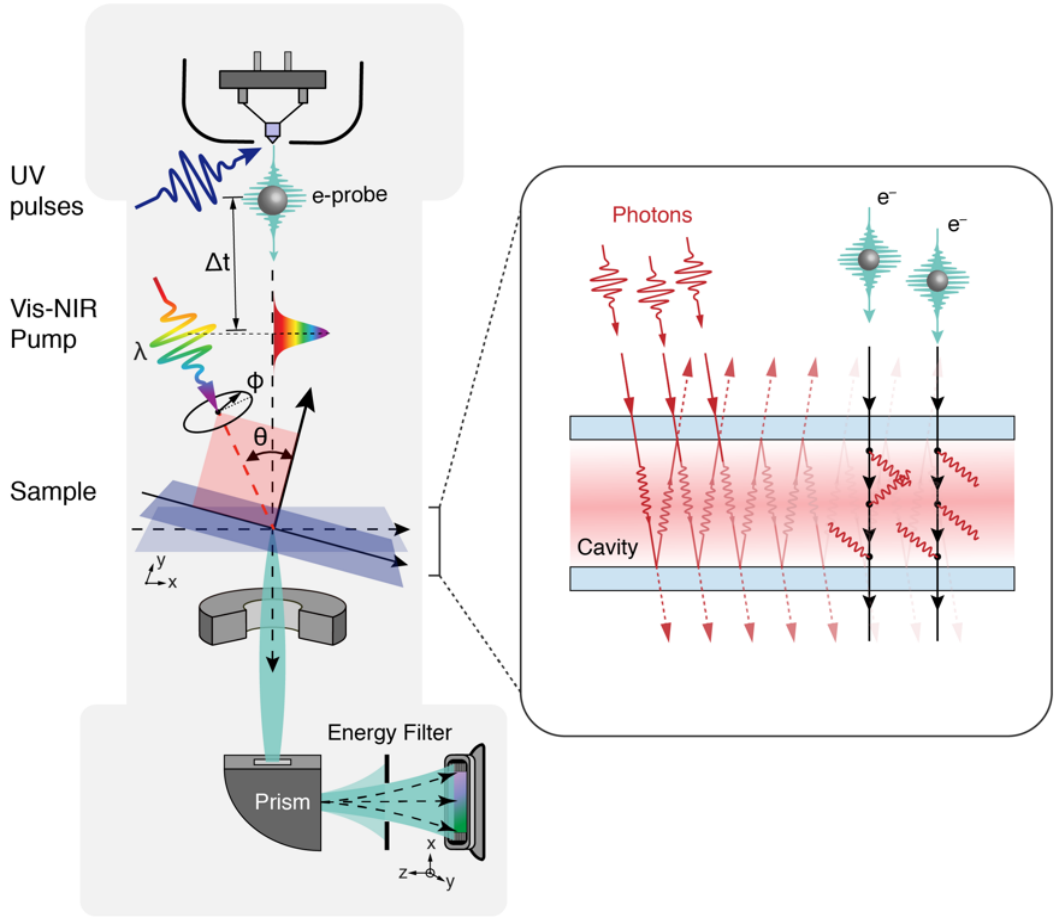

**Figure 1 | Cavity quantum electrodynamics (CQED) with free electrons in the ultrafast transmission electron microscope.** Schematic of the setup, offering five degrees of freedom: the delay $\Delta t$ between the light pump and electron probe, the pump light wavelength $\lambda$ and polarization $\varphi$, sample tilting angle $\theta$, as well as the electron spatial distribution in the $xy$ plane after the interaction (elaborated in Fig. 2). We can image the electrons with or without energy-filtering. Inset: schematic image showing the quantized interaction of free electrons with cavity photons.

**Results**

The enabling technique in our work is that we probe an optical cavity (specifically a two-dimensional photonic crystal membrane) with multidimensional capabilities that we developed in the ultrafast transmission electron microscope (Fig. 1). As in other such microscopes, a femtosecond laser pulse is split into two parts, one used to excite (pump) the sample and the second to generate the electron pulse that interacts with (probes) the sample. The probe electron is used to image the optical field distribution in real space by electron energy filtering (Fig. 2a), resolving the nearfield of nanostructures at deep-subwavelength resolution. In our setup, we can also vary the sample tilt relative to the laser pump (Fig. 2b) to enable coupling to photonic cavity resonances. The resonances can be fully characterized by changing the pump laser wavelength (Fig. 2c) and polarization (Fig. 2d). Using a controllable delay time between the pump laser pulse and probe electron pulse (Fig. 2e), we can image the dynamics of light confined in the photonic cavity at ultrafast timescales. More information about the setup is provided in the Methods section.

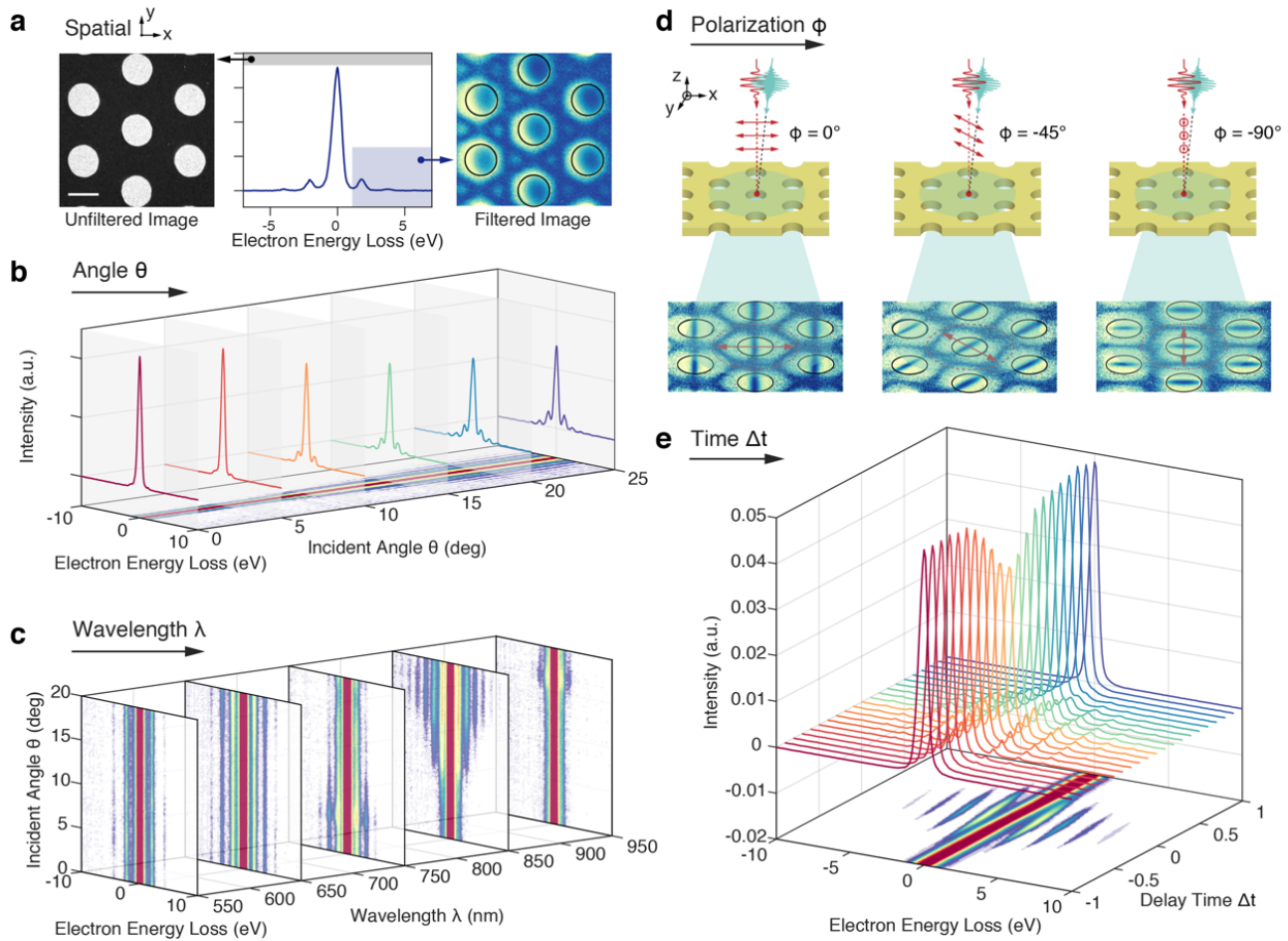

**Figure 2 | Ultrafast transmission electron microscopy for multidimensional spectroscopy:** space, momentum, energy, polarization and time. **a,** Electron microscopy images of a photonic crystal membrane used for proof-of-concept demonstrations throughout this work. Imaging matter/light-field with the electron energy filter disabled/enabled (left/right respectively). The middle plot shows a typical electron energy loss spectrum (EELS); energy-filtered electrons used for imaging are marked by blue shaded area. Inset white scale bar is 300 nm. **b,** Multiple EELS over a range of sample tilt angles, showing the angle-resolved capability; the bottom plane displays an angle-resolved EELS map assembled from individual spectra. **c,** EELS map showing the capability of mapping the electron–photon interaction for a range of wavelengths. **d,** Snapshots showing the capability to image excitations of different incident light polarizations. **e,** Multiple EELS showing the femtosecond time resolution of the interaction; the bottom plane displays a time-resolved EELS map assembled from individual spectra.

To demonstrate the above capabilities and investigate the strong electron–cavity photon interaction, we characterized the interaction of free electrons with cavity photons created in a triangular photonic crystal. First, we used the ultrafast transmission electron microscopy to measure the bandstructure of the photonic crystal (Fig. 3a) for both TM and TE polarizations in the wavelength range 525 to 950 nm and incident angles from 0° to 24.4° (See Methods). Our spectral resolution was limited only by the laser linewidth of 5–10 nm in the visible range and by our choice of a 5 nm wavelength step size. Our angular resolution was limited by our choice of the stage tilting step of 0.2°. The measured bandstructure (Fig. 3a) agrees very well with the simulated one (Fig. 3b), calculated by the finite-difference time domain (FDTD) method (Supplementary Note 1). The measurement and simulation are performed along the Γ−K axis, as defined in Fig. 3c and d. The slight artifacts in the measurement appeared because of a small polarization impurity in our setup (Supplementary Note 2).

In addition, for every point measured in the bandstructure, we were able to extract the spatial distribution of the electric field (typical examples are shown in Fig. 3e and Fig. 2a; see Supplementary Note 3), which is directly measured by the electron interaction with the electric field component parallel to the electron velocity[9–11]. This interaction enables the direct imaging of the photonic crystal Bloch mode at deep-subwavelength resolution – ~30 nm (Fig. 3d 10). Photonic Bloch modes were previously imaged in photonic lattices[31,32] above the diffraction limit, or by observing the local density of states with nearfield probes on photonic crystal waveguides[33,34] and slabs[35,36]. Our measurement was not performed only at either TM or TE, but also in their superposition states, demonstrating the rotation of the photonic crystal Bloch mode's spatial distribution (shown in Fig. 3f, using 730 nm wavelength and normal incident laser excitation).

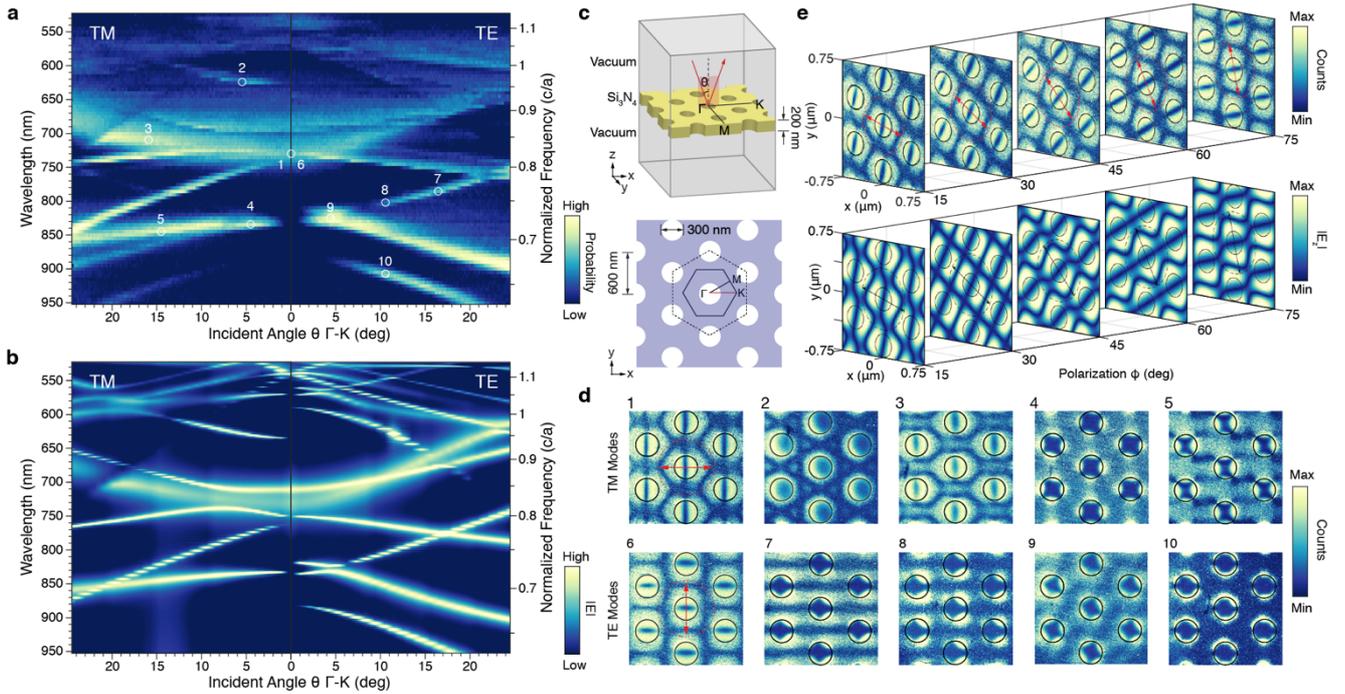

**Figure 3 | Bandstructure reconstruction and direct imaging of photonic crystal Bloch modes. a,** Measured bandstructure produced by scanning over laser incident angles and wavelengths. Each datapoint in the map is a separate electron energy loss spectrum (EELS) measurement of the electron–light interaction at zero delay time. Color scales as the interaction probability. **b,** Simulated bandstructure by the finite difference time domain (FDTD) method – see Supplemental Note 1. **c,** Diagram showing the layout of the photonic crystal and incident pump laser pulse. **d,** Photonic crystal Bloch mode images as measured in the ultrafast transmission electron microscope for different angles and wavelengths (marked in the bandstructure). **e,** Measured (top) and FDTD simulated (bottom) photonic crystal Bloch mode images rotating with the polarization direction (indicated by double arrows) at normal incidence and 730 nm wavelength (points 1 and 6 in the bandstructure).

In addition to the full spatial and energy–momentum distributions of the photonic crystal membrane, we measured the photon lifetime of a cavity mode directly *inside* the cavity by an electron probe (Fig. 4). This feat was made possible by the ultrafast temporal resolution of our system and the relatively high quality factor (Q) achievable in photonic crystal modes. Fig. 4a compares the electron energy loss spectra (EELS) achieved by interacting the electron probe with a high or a low-Q photonic crystal mode, presented as a function of the delay between the laser and electron pulses. Depending on the lifetime of the photon in the cavity, the spectrum exhibits typical time-symmetric behavior (for low-Q) or time-asymmetric behavior (for high-Q) around zero delay time. Note that the slight time–energy tilt of both spectra in Fig. 4a and b is a result of the electron dispersion[37].

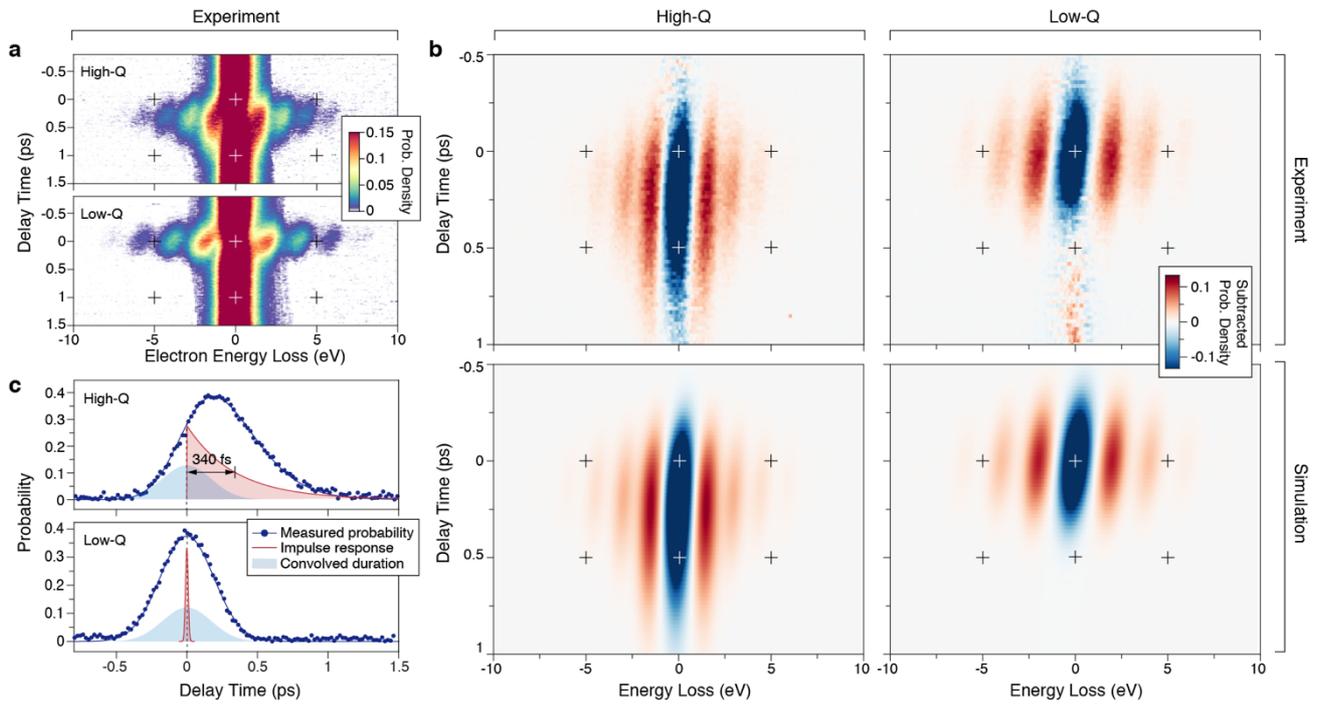

**Figure 4 | Direct measurement of the photon lifetime and dynamics by a free electron. a,** Time-resolved electron energy loss spectra (EELS) maps obtained from a high-Q mode (point 4 in Fig. 3) and a low-Q mode (point 2 in Fig. 3). The crosses help to identify the zero energy loss and zero delay time. The color represents the probability of finding an electron at each energy, with the strong zero-loss peak in dark red. **b,** Time-resolved difference EELS map after subtracting the zero-loss electrons, showing the difference between high-Q and low-Q modes. The high-Q map (upper left) is elongated and time-asymmetric in contrast with the shorter and time-symmetric low-Q map (upper right). Using Eq. 1, the results of simulations accounting for (lower left) and neglecting the cavity photon lifetime (lower right) match the experimental results. **c,** Experimental (dots) and simulated (solid line) probability density of the electron–photon interaction. The light-blue area indicates the combined durations of the electron and laser pulses, highlighting the asymmetric features (top panel). The impulse response of the high-Q and the low-Q modes are presented in red, showing an exponential decay with lifetime of 340 fs, and a shorter lifetime below our measurement resolution (effectively a Dirac Delta function).

To extract the photon dynamics and lifetime, we separated the interaction duration from the background of the non-interacting electrons (see Methods), also eliminating various noise factors, such as camera dark noise. The extracted dynamics are shown in Fig. 3b, where the interaction time difference is quite visible: the high-Q spectrum is elongated, whereas the low-Q spectrum persists only during the overlap of the pump and probe pulses. By fitting the experimental results in Fig. 4b to theory (lower panels in Fig. 4b), we found the photon lifetime $\tau$ of the high-Q spectrum to be approximately 340 fs and its quality factor ($Q = \pi \tau c/\lambda$) to be

~384. We further corroborated these Q values using the spectral linewidths extracted from numerical simulations (Supplemental Note 4). The high Q value implies more than two orders of magnitude *enhancement* in the probability of interaction with the trapped photon and later showing more than an order of magnitude enhancement in the interaction strength relative to the current record with a metal mirror. We also confirmed the persistence of light in the cavity after the initial excitation pulse by comparing the interaction time to the convolution of the durations of the electron and laser pulses (Fig. 4c). The results show that the maximal point of interaction is shifted by ~207 fs, which further confirms that the total optical energy in the mode is increased over time.

The theory describing this phenomenon requires the addition of the photon lifetime to the conventional photon-induced near-field electron microscopy (PINEM) theory[10,11]. The probability density per unit energy that an electron absorbs/emits $l$ photon in a cavity with lifetime $\tau$ is

$$P_l(E,t) = J_l^2\left((\Theta(t)e^{-\frac{t}{\tau}}/\tau) * |2\beta(t)|\right) * G(t - \zeta E, \sigma_e), \tag{1}$$

where $*$ denotes convolution, $E$ is electron energy, $t$ is the delay time, $J_l$ is the $l$th order Bessel function of the first kind, and $\Theta(t)$ is a Heaviside step function. $\beta(t) = \beta_0 e^{-(t/\sigma_L)^2}$ is the time-dependent PINEM field[10] that quantifies the strength of interaction with the laser pulse by a dimensionless parameter $\beta_0$. $G(t,\sigma) = \frac{1}{\sigma\sqrt{\pi}}e^{-(t/\sigma)^2}$ describes the electron pulse duration, into which we substitute the intrinsic chirp coefficient $\zeta$. The standard deviations $\sigma_e, \sigma_L$ respectively of the electron and the laser pulses depend on the full-width half maximum (FWHM) durations $\tau_e, \tau_L$ via $\sigma_{e,L} = \tau_{e,L}/(2\sqrt{\ln 2})$. Note that in the limit $\tau \to 0$, Eq. 1 converges back to the conventional theory[10,11]. See Supplementary Note 5 for more details on using Eq. 1 to fit the experimental results to theory, as shown in Fig. 4b.

The strong cavity enhancement enables record-low laser pulse energies to be used for electron–photon interactions. Fig. 5a shows a comparison of the interaction energy spectra, measured with the same laser pulse excitation (730 nm wavelength, 1 nJ pulse energy at normal incidence), for both a photonic crystal membrane and a thin metallic film. The excitation wavelength was selected to maximize the confined electric field inside the photonic crystal, greatly increasing the probability for multiphoton interactions as compared to a standard interaction platform, e.g., a metallic film[38]. Moreover, in Fig. 5b, the dependence of the electron–cavity photon interaction probability on the incident laser pulse energy is shown, confirming that an interaction remains visible for pulse energies as low as 100 pJ (2.67 µJ/cm² fluence). The inset of Fig. 5b also shows the total interaction probability as a function of pulse energy, revealing a cavity enhancement of more than an order of magnitude as compared to the metallic film.

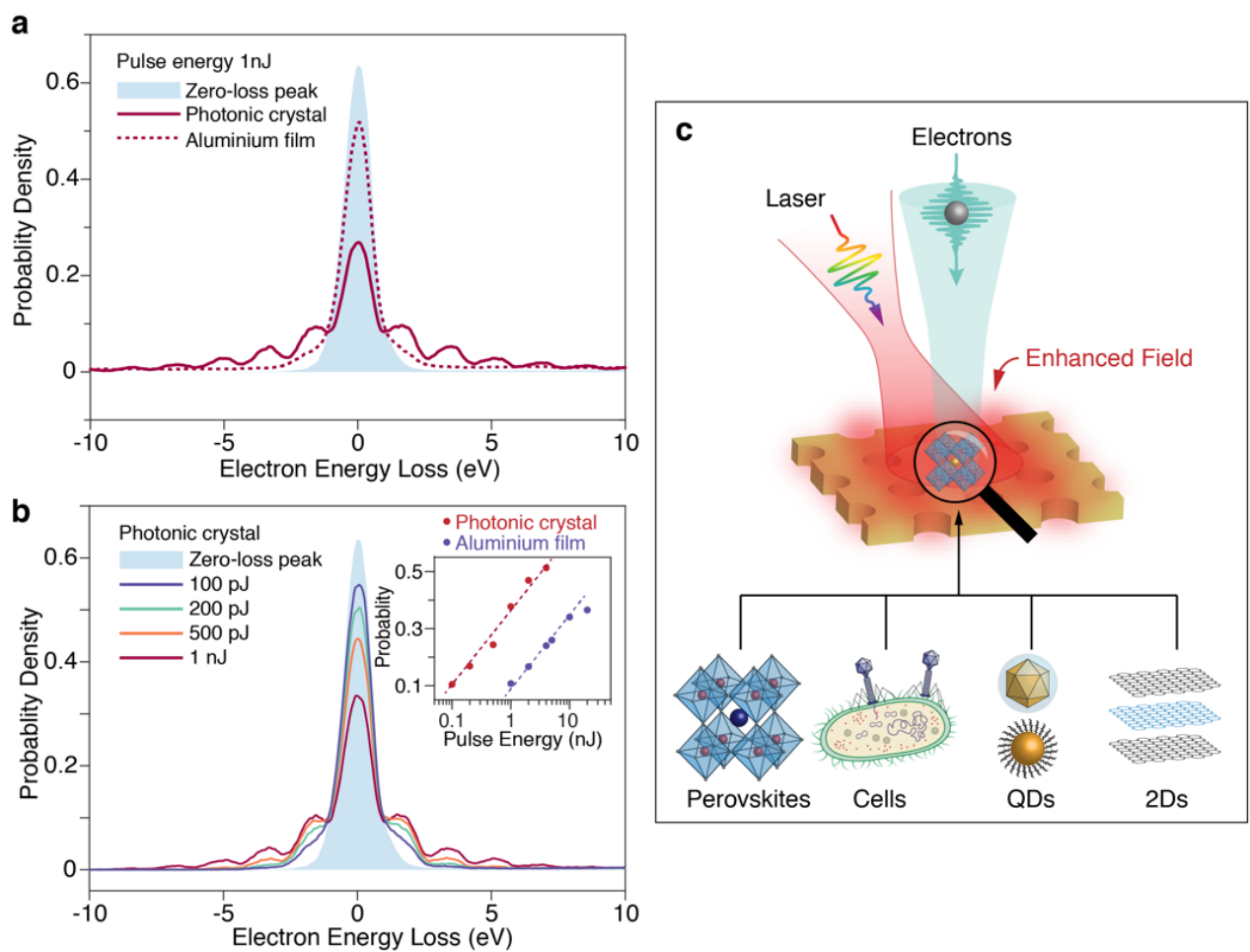

**Figure 5 | Enhanced interaction of electrons with cavity photons achieving record-low pulse energy. a,** electron energy loss spectra (EELS) measured with 1 nJ pulse energy excited at 730 nm wavelength and normal incidence. Comparison of non-interacting electron spectrum (blue shaded area) with spectra of electrons interacting with a photonic crystal (solid line) and a 31 nm aluminum film (dashed line), showing a significant interaction enhancement. **b,** EELS of the interaction with the photonic crystal mode for different laser pulse energies, showing that the interaction persists even at energies as low as 100 pJ. The inset presents the total interaction probability as a function of laser pulse energy, showing an order of magnitude decrease in the required energy, between the aluminum sample and photonic crystal membrane. This comparison *demonstrates the large enhancement* to the electron–cavity photon interaction. 1 nJ corresponds to 26.7 µJ/cm$^2$ fluence (the laser spot diameter was ~69.0 µm in this figure, see Supplemental Note 6). **c,** Prospects of photonic cavity structures as platforms for low-dose excitations of beam-sensitive samples for ultrafast multidimensional spectroscopy and microscopy.

**Discussion**

In summary, we directly measured the lifetime of cavity photons via a free electron probe and achieved coherent electron–photon interaction at record-low pulse energies. Simultaneously, we were able to record the complete real space and energy–momentum space information of our sample. Our work will help promote additional important capabilities of nearfield imaging in ultrafast transmission electron microscope, such as imaging nearfields residing deep inside materials[39] and extending outside them[12], without the probe introducing near-field distortions[40]. In particular, our measurement of the field in the holes of a photonic crystal illustrates the potential of probing light inside hollow structures (nanotubes, hollow-core fibers, and dielectric laser accelerators[41–43].

Our method combines simultaneously the characterization abilities of different nearfield setups, in space[33–35,40] and time[40,44–48], energy–momentum space[36,49,50], and polarization[36,51], as well as combinations thereof[35,36,45,51,52]. In this respect, our work[20] has been developed in parallel with other efforts[53,54] to pursue the full integration of all of the above capabilities, at a comparable or better resolution, in a single setup.

The significant enhancement of electron–photon interaction by using cavity modes suggests a path toward low-dose excitation of soft matter and other beam-sensitive samples (e.g., halide perovskites). One could place a delicate sample, biological or otherwise, on an optical cavity (Fig. 5c) to enable ultrafast multidimensional spectroscopy and microscopy using lower pulse fluence to reduce sample damage. As such, the method presented in this work can be readily integrated[55] with established techniques, such as cryogenic electron microscopy[56] or liquid-cell electron microscopy[57]. The low excitation dose is achieved by virtue of the field enhancement resulting from the cavity resonance and the small number of electrons in the ultrafast electron microscopy experiments[9].

The exploration of higher-Q cavity modes in our system is limited by the laser pulse's coupling to the cavity, which depends on their linewidth overlap. Looking ahead, the coupling could be greatly improved by using nanosecond[58] or even continuous wave lasers[59] that have considerably narrower linewidths. Such lasers could efficiently couple to cavities with ultrahigh-Q (e.g., ~$10^6$ demonstrated in extended cavities with bound states in the continuum[60], and ~$10^8$ in ring cavities with whispering gallery modes). Efficient coupling to these ultrahigh-Q cavity modes inside electron microscopes could lead to new types of coherent electron–photon interactions, such as novel opto-electro-mechanical effects[61]. Such cavities enable strong coupling of light with mechanical vibrations[61,62], which can now be combined with the prolonged coherent electron–cavity photon interaction (as demonstrated in this study). By further exploiting the influence of mechanical vibrations on the output phase of electrons passing through the resonator, a unique nonlinear interaction of laser pulses, mechanical vibrations, and free electrons may be achieved[61].

The high multidimensional resolution of our method may also be used to explore electromagnetic structures at the nano- or even pico-scale, the characterization of which is limited in other methods. For example, current methods for exploring extreme nanophotonic structures, such as pico-cavities[22], frequently involve undesired collective effects and background signal from multiple structures that cannot be easily separated because of the limits of optical resolution (e.g., incoherent broadening). Our method can vastly contribute to the study of these emerging systems, as it enables the full investigation of a single cavity at a time, potentially unveiling the mysteries associated with the dynamics of the single atoms therein[22]. A further example could be the in situ characterization, via the ultrafast electron probe, of topological photonic mode dynamics, associated with novel opto-electronic devices[63,64].

Although this work was focused on exciting *optical* modes in nanostructures, the ability to excite optically *electronic* systems, such as quantum dots[21] and van der Waals heterostructures[23,24], suggests an alternative probing mechanism. The optical excitation could create out-of-equilibrium initial conditions in the material that is subsequently probed with the electron pulse. Such a method would resemble two-photon angle-resolved photoemission spectroscopy[65], but would not be limited by the momentum and energy of the probe photon in terms of supplying a full electronic characterization in real-space, energy-momentum space, and time.

Finally, our work promotes the inclusion of free electrons in the established field of CQED, which has thus far been focused only on bound electron systems. Similarly to a two-level atom in a cavity that exhibits vacuum Rabi oscillations, a free electron can absorb a photon from vacuum and re-emit it into the cavity several times, depending on the electron coherent pulse length and the cavity quality factor. In this framework, strong coupling between free electrons and cavity photons may be achievable[25,26]. Electron–photon cavity interaction could also be used to manipulate the electron quantum state, suggesting the use of free electrons as a qubit[66] or as a carrier for transferring quantum information.

**Methods**

*Ultrafast electron microscopy:* The experiments were performed on an ultrafast transmission electron microscope that is based on a JEOL 2100 Plus transmission electron microscope (TEM) with an $LaB_6$ electron gun (acceleration voltage varies from 40 kV to 200 kV), the schematic of which is shown in Fig. 1. The ultrafast electron transmission microscope is a pump-probe setup that uses femtosecond light pulses for exciting the sample and ultrafast electron pulses for probing the sample's transient state. To this end, a 1030 nm, ~220 fs laser (Carbide, Light Conversion) operating at 1 MHz repetition rate is split into two pulses. The first pulse is converted to UV via two stages of second-harmonic generation and then guided to the TEM cathode by an aluminum mirror inserted in the TEM column. This process generates ultrafast electron pulses. These electron pulses travel along the z-axis, penetrate the sample and image it. The second pulse is converted into variable wavelengths by an optical parametric amplifier (OPA) for pumping the sample. This pulse is finally guided by an additional aluminum mirror in the TEM column and incidents onto the sample from the top with a small angle ~4.4° relative to the *z*-axis in the *xz* plane. The delay time between the electron pulse and OPA pulse is controlled by a motorized stage. The photonic crystal sample (Ted Pella, Pelco #21588-10) is installed on a double-tilt TEM sample holder that allows tilting around the *x* and *y* axes from -20° to 20°. To analyze the electron energy spectrum after interaction, a post-column electron energy loss spectroscopy system (Gatan) is installed in the TEM. This system also provides the energy-filtered TEM capability using the EELS system for real-space imaging. The inclusion of all the above multidimensional capabilities in one setup is extremely useful for full characterization of nano-scale objects, e.g., alleviating risks of losing the region of interest during the transfer of the sample between setups.

*Bandstructure reconstruction:* For mapping the bandstructure, we operated the ultrafast transmission electron microscope in TEM mode at 80 keV electron energy and parallel illumination. The EELS are collected over a range of wavelengths from 525 nm to 950 nm and incident angles from 0° to 24.4° with a zero-loss peak ~1.1 eV. The measured PINEM spectra are centered and normalized to probability one to reduce noise from fluctuations in the electron current. Then, the probability of the electron interaction with the optical nearfield is calculated by integrating the electron energy spectra outside a range that is twice the zero-loss peak FWHM. The details of this data processing are provided in Supplementary Note 7.

*Optical nearfield imaging:* We used energy-filtered TEM at 200 keV to image the light field with deep-subwavelength resolution, while providing sufficient electrons that penetrate the $Si_3N_4$ membrane. The images are acquired in energy-filtered mode with a slit in the energy spectrum that has a width of ~10 eV and is centered at -10 eV (energy gain side). To reduce the contribution of scattered electrons, an objective aperture is applied during image exposure. We find an approximately 87.5% count loss for electrons that penetrate the $Si_3N_4$ membrane, as compared to electrons that move through the holes. To show the light field nanostructure in the membrane more clearly, post image processing is introduced to enhance the contrast of the image in the membrane area. Consequently, the signal-to-noise ratio is lower in the membrane area. (See Supplementary Note 8 for more details).

*Cavity photon lifetime and field enhancement:* The EELS are collected as a function of delay time with 200 keV electrons in TEM mode. We measured the reference zero-loss peak by probing the photonic crystal sample with the electron pulse a few picoseconds before the laser excitation. This reference zero-loss peak is used as the background of the non-interacting electrons in the time-resolved EELS map (Fig. 4a). After subtracting the zero-loss electrons, the time-resolved difference map was obtained (Fig. 4b). The field enhancement (Fig. 5) is measured by comparing electron interactions with excitations of different modes of the photonic crystal, in addition to laser excitation of an evaporated aluminum film (Ted Pella, Pelco Product #619), used as a reference sample.

**Acknowledgements**

K.P.W. is supported in part at the Technion by a fellowship from the Lady Davis Foundation. I. K. acknowledges the support of the Azrieli Faculty Fellowship. The experiments were performed on the ultrafast transmission electron microscope of the I. K. AdQuanta group installed in the electron microscopy center (MIKA) in the Department of Material Science and Engineering at the Technion.